\documentstyle[12pt,epsfig]{article}
\begin{document}
\begin{center}

\baselineskip=24pt

{\Large \bf Narrow muon bundles from muon pair production
in rock}

\vspace{1cm}

{\large V. A. Kudryavtsev$^a$, E. V. Korolkova$^b$ and N. J. C. Spooner$^a$}
\vspace{6pt}

\baselineskip=18pt
\vspace{0.5cm}

$^a${\it Department of Physics and Astronomy, University of Sheffield, 
Sheffield S3 7RH, UK}

$^b${\it Institute for Nuclear Research of the Russian Academy of Science, 
Moscow 117312, Russia}

\vspace{1cm}
{\large \bf Abstract}

\end{center}

We revise the process of muon pair production by
high-energy muons in rock using the recently published cross-section.
The three-dimensional Monte Carlo code MUSIC has been used to obtain
the characteristics of the muon bundles initiated via this process. We
have compared them with those of conventional muon bundles initiated in
the atmosphere and shown that large underground detectors, capable of
collecting hundreds of thousands of multiple muon events, can discriminate
statistically muon induced bundles from conventional ones. However, 
we find that the
enhancement of the measured muon decoherence function over that predicted
at small distances, recently reported by the MACRO experiment, cannot be
explained by the effect of muon pair production alone, unless its cross-section
is underestimated by a factor of 3.

\vspace{1.0cm}


\vspace{1cm}

\noindent Corresponding author: V. A. Kudryavtsev, Department of Physics and 
Astronomy, University 
of Sheffield, Hicks Building, Hounsfield Rd., Sheffield S3 7RH, UK

\noindent Tel: +44 (0)114 2224531; Fax: +44 (0)114 2728079

\noindent E-mail: v.kudryavtsev@sheffield.ac.uk

\pagebreak

\noindent {\large \bf 1. Introduction}
\vspace{0.5cm}

Muon bundles, or events with muon multiplicity more than 1, are studied
in underground experiments to obtain information on the primary
cosmic-ray composition and characteristics of hadron-nucleus
interactions. Conventional techniques include comparison of the
measured multiplicity and pair separation distributions with
predictions based on primary composition models and
models of the development of hadronic cascades in the atmosphere.
There is, however, an effect which can slightly modify such distributions.
This is the process of muon pair production by muons in the rock (or water)
above the detector. The cross-section of the process is quite small but its
effect is not negligible for small
separation distances between muons in the bundles. 
Double or triple muon events observed in underground experiments can arise
from either multiple muon production in Extensive Air Showers (EAS)
in the atmosphere or muon pair production by single muons in the rock
(a double muon event is observed if one of the muons is stopped on its way
to the detector). As single muons dominate over multiple muons 
underground, muon bundles produced by single muons in the rock can 
contribute to the total number of bundles. An excess in
the number of detected bundles over the predictions of EAS models can then
be visible.
Muon bundles produced in water are the background in underwater 
detectors looking for up-going neutrino-induced muons 
\cite{AMANDA,Baikal,ANTARES,NESTOR}.

Original estimates of the fluxes of muon bundles from 
muon pair production using one-dimensional calculations have been published
in \cite{KKL}. First
three-dimensional simulations of the muon transport through standard rock
taking into account muon pair production have been performed recently 
\cite{VAK1}. In both cases authors used the cross-section 
calculated in \cite{Kelner} and parametrised in \cite{Bugaev}. In 
\cite{VAK1} it was shown
that the process should contribute to the number of narrow muon bundles 
detected in underground experiments, such as LVD \cite{LVD} and 
MACRO \cite{MACRO}. The observation of this effect has been recently
reported by the MACRO collaboration \cite{MACRO1}.
However, the cross-section from \cite{Kelner,Bugaev}
used in \cite{VAK1} was obtained assuming
point-like nuclei with complete screening, which is not a reasonable
approximation in this case (see \cite{KKP} for the discussion).
Since then, a new, more accurate cross-section for muon pair production
by muons has been obtained \cite{KKP}. The cross-section calculated in 
\cite{KKP} is roughly 2-5 times smaller than that of
\cite{Kelner,Bugaev}. The new cross-section has been used by the authors 
of \cite{KKP} to calculate
total fluxes of muon bundles produced in the rock for different depths 
of observation underground using an one-dimensional numerical 
integration technique. 
However, the total fluxes of muon-induced bundles are hidden by the
two-orders of magnitude higher flux of
conventional muon bundles initiated in EAS in the
atmosphere. Muon bundles from muon pair production
can be discriminated only by the characteristic
that they have small separation between muons in the bundle \cite{VAK1}
(later on we will call them narrow muon bundles). Small muon pair
separation in such events arises because they are produced
in the rock quite near the observation level, while conventional
muon bundles are produced in the upper layers of the atmosphere.
Thus, a full three-dimensional
Monte Carlo simulation is necessary to evaluate the effect caused
by muon pair production and to give predictions for underground experiments.

In this paper we describe such simulations using the new cross-section
\cite{KKP} and discuss the implications of the results
for underground experiments. We show that the excess of narrow muon bundles,
caused by their production in rock, over predictions from
EAS models is visible and can be measured experimentally. Such measurements
performed at various depths underground could be used as a test of the
cross-section at different energies.
Our calculations differ from those in \cite{VAK1} by the use of the new,
more accurate, cross-section of muon pair production by muons from 
\cite{KKP}. They differ from the semi-analytical approach in \cite{KKP} 
by the use
of a full three-dimensional Monte Carlo technique which is the only way to
give predictions for muon separation in the bundles.

\vspace{0.8cm}
\noindent {\large \bf 2. Muon transport through rock and water}
\vspace{0.5cm}

The three-dimensional Monte-Carlo code MUSIC \cite{MUSIC}
is designed to simulate 
propagation of muons from sea level down to the level of observation.
The code takes into account the stochasticity of all processes 
with fractional energy transfer $v>10^{-3}$.
It calculates the angular deviation and lateral displacement of muons
due to multiple scattering and sto\-chastic processes. The code is
regularly updated \cite{VAK3} and includes the most recent and accurate
cross-sections of muon interactions in matter: bremsstrahlung
\cite{KKP1}, electron-positron pair production \cite{KKP2} (with
the corresponding cross-section on atomic electrons from \cite{Kelner1}),
and inelastic scattering \cite{BB}.

The latest version of the code has been modified to
take into account muon pair production with 
fractional energy transfer $v>10^{-3}$ with the
cross-section from \cite{KKP}. 
An analytical expression for the differential cross-section as a function 
of final muon energies was obtained \cite{KKP}
by analogy with similar cross-section for electron pair production by muons
\cite{KKP2}. Comparison with precise numerical integration
done in \cite{KKP} showed that the accuracy of 
parametrization is better than 10\% for initial muon energy
greater than 10 GeV and final muon energies greater than 1 GeV, while
the accuracy of the
total cross-section is better than 3\% at initial muon energies
greater than 30 GeV. Effects of nucleus screening and finite nucleus size
were taken into account in the calculations \cite{KKP}. 

We have not restricted the simulation
to only one interaction with muon pair production (as had been done in 
\cite{VAK1}), but considered
all possible interactions which, in principle, can result in a muon bundle
with multiplicity more than 3.

The differential cross-section as a function of the scattering angle of
each muon is not available. Therefore,
the scattering of muons at the point of muon pair 
production has not been taken into account.
However, in general, the scattering of muons due to stochastic processes
can be neglected, because the resulting angular deviation and lateral
displacement are fully determined by the multiple Coulomb 
scattering \cite{MUSIC}.

We want to emphasize also that precise simulation of the muon bundles
initiated in EAS in the atmosphere is out of the aims of this paper.
To estimate the significance of the effect caused by muon pair production
in rock, we will compare the calculated flux of muon bundles from this process
with the flux of muon bundles measured in underground experiments.
To calculate the flux of muon-induced bundles it is enough to use a power-law
parametrization of muon energy spectrum at the surface without detailed 
knowledge of muon multiplicity and separation in a particular shower. 
This is because single muons dominate over multiple muons by almost one 
order of magnitude (for existing underground detectors) and, hence, single 
muons will provide major contribution to the effect mentioned above.

For comparison we have carried out the simulations both in standard
rock and in water.
$10^7$ single muons with an initial energy at sea level sampled 
according to a
power-law spectrum with power index -3.7, were propagated down to
different depths in rock and water. A lower edge of the spectrum at sea level
was chosen for each depth separately to ensure a muon survival
probability $\le 0.003$. This means that all muons which have a probability
of more than 0.003 of reaching a predefined depth, have been included in the
analysis. Note that the muon energy spectrum at sea level has a more
complicated form than a simple power-law at energies below 1 TeV and, hence, the
results for shallow depths may be biased. However, the purpose of this work
is not to provide exact estimates of this effect, which is dependent
on the details of individual experiments, but only to evaluate
the significance of the effect and to show how it can be measured.

\vspace{0.8cm}
\noindent {\large \bf 3. Results and discussion}
\vspace{0.5cm}

The results of the simulations are presented in Table 1.
Mean survival probabilities for single muons 
($N_{surv}/N_{sim}$ in Table 1) agree 
well with earlier calculations
\cite{VAK1} and with the results of the original code \cite{MUSIC}.
The ratios of the total numbers of narrow muon bundles to that of single muons
($R_2$, $R_3$ and $R_b$ in Table 1) are significantly
smaller (by a factor of approximately 3 at 3 km w.e.) 
than those obtained in \cite{VAK1} due to the different muon pair production
cross-section used but agree well with semi-analytical calculations
\cite{KKP} (see also Figure 1 for comparison with the results from \cite{KKP}). 
Table 1 shows also the main characteristics for narrow bundles 
and single muons, such as:
mean energy of single muons $<E_{s}>$,
mean energy of muons in bundles $<E_{b}>$,
mean path of muon bundle in the rock/water between the point of its production
and the observation level $<L>$,
mean muon pair separation in double ($<D_2>$), triple
($<D_3>$), and double + triple ($<D_b>$) muon events,
mean angular separation of muons in bundles $<\alpha>$, 
and the ratio of
number of pairs with separation less than 1 m to the number
of single muons $R_b(<1$ m). 
The last column shows the results for 3 km of water. 
To calculate mean pair separation,
mean angular separation of muons and the number of muon pairs we applied 
weighting
$1/N_p$, where $N_p$ is the number of independent muon pairs in the bundle:
$N_p=1$ for double muon events and $N_p=3$ for triple muon events. This has
been done to allow easy comparison between our simulations and 
results of studies of muon bundles detected in underground experiments
(see, for example, \cite{LVD,MACRO1}).
Only muons with energy more than 1 GeV (a typical threshold in underground
experiments) have been taken into account.
The ratios of the number of muons bundles to that of single muons together
with the results of \cite{KKP} are shown also in Figure 1.

As can be seen in Table 1 and Figure 1, the ratios of narrow muon bundles 
to single muons ($R_2$, $R_3$ and $R_b$)
increase with depth. This results from the rise in interaction 
cross-section with muon energy. As the mean muon energy which contributes to
the muon flux at a particular depth increases with depth, the effective
cross-section also increases and hence so does the fraction of narrow muon
bundles with respect to single muons. The mean energy of muons in bundles
($<E_{b}>$)
is roughly twice that of single muons ($<E_{s}>$) 
and increases with depth. 
Muons in bundles cross on average 1/5 of the total thickness of rock
from the point of interaction down to the observation level.
The ratio of double to triple muon events ($R_2/R_3$) 
increases with depth due to the
increase in thickness of rock crossed by the bundle ($<L>$) and, hence, 
due to the decrease in the probability that all three muons survive.
The increase in mean muon energy in 
the bundles partly compensates for this effect.

The weighted mean separation of muon pairs ($<D_2>$, $<D_3>$, $<D_b>$)
increases by a factor of 2 -- 2.5 with increase of depth from 1 to 
10 km w.e., while the weighted mean angular separation for pairs 
($<\alpha>$) decreases by a
similar factor. This can be explained in the following way. Pair separation,
determined in this case by multiple Coulomb scattering, depends mainly on
the muon path in the rock. As the mean path of muons in bundles down
to the observation level ($<L>$) increases with depth, pair separation also
increases. 
The increase in mean muon energy partly compensates for the rise in ($<L>$).
The angular separation of muons due to Coulomb scattering 
is determined mainly by the muon energy. Since mean
muon energy increases with depth, mean angular separation decreases.

We have found the mean separation of muon pairs to be about $50\%$
higher than obtained in \cite{VAK1}. This is possibly explained by the
larger thickness of rock crossed on average by muon bundles in the 
case of a smaller bundle production 
cross-section (see also the comparison between values
of $<L>$ in rock and water in Table 1).

The ratio of muon bundles to single muons in water is smaller than that
in the rock. Such behaviour is expected from the $Z^2/A$ - dependence
of the macroscopic cross-section on the target nucleus charge and mass.
Weighted distributions of pair separation distances for a depth of
3 km w.e. in rock and water are presented in Figure 2. 
Due to the lower density of water, the distribution in water is wider
than that in the rock.

The small mean pair separation of these bundles can be used to discriminate
them statistically from conventional muon bundles originating 
in Extensive Air Showers in the atmosphere. The ratio of conventional 
muon bundles to single muons is typically $(1-10)\%$
depending on the detector geometry. The mean pair
separation of conventional muon bundles is of the order of several metres.
Only a few percent of muon pairs in conventional muon bundles contribute
to separation distances less than 1 m, while
from $60\%$ to $90\%$ of pairs in narrow muon bundles have muons
separated by less than 1 m.
The ratio of the weighted number of pairs with separation distances less than
1 m in narrow bundles to the number of single muons 
($R_b(<1$ m) in Table 1, see also Figures 1)
is $2.5 \cdot 10^{-4}$ at 3 km w.e. This number has to be compared with
$\approx 2 \cdot 10^{-3}$, the estimated ratio of pairs with
separation distances less than 1 m in detected bundles to single muons 
\cite{VAK1}. A $10\%$ effect can easily be found if the total statistics 
contains hundreds of thousands of muon bundles. This is certainly within 
the reach of modern underground detectors. 

It is of interest to consider the LEP detectors at CERN \cite{L3,Aleph}.
Despite being located at shallow depth (where the ratio of narrow muon bundles
to single muons is quite small), these detectors, having good
spatial, angular and energy resolution and
the ability to collect millions or even billions of muon events
\cite{L3}, could be a means of searching for the effect above.

Existing deep underground experiments such as MACRO \cite{MACRO} 
and LVD \cite{LVD}
are also able to search for an excess of narrow muon bundles over the
expected number of conventional muon bundles. In fact, evidence for
an excess at separation distances less than 3 m has recently been
published by the MACRO collaboration \cite{MACRO1}. The authors explained
the excess as due to production of muon pairs in the rock. However, to
calculate the predicted rate they used the cross-section from 
\cite{Bugaev}, which gave an overestimate of the effect by a factor of 3.
This means that their measured rate should be about factor of 3 higher than
expected with the newly calculated cross-section from \cite{KKP}.
Moreover, the measured pair separation distribution (after subtraction
of the contribution from conventional muon bundles) is much wider than 
expected from our simulations. Note, that their calculated mean
separation distance is almost twice that of our result for 3 km w.e.,
which is hard to explain by the difference in the cross-sections used and
the contribution from muons crossing larger thicknesses of rock.
Our simple estimates, based on the published pair separation distribution
\cite{MACRO1} and relative rates of single and multiple muon events
\cite{MACRO}, show that MACRO observed the ratio of weighted number of
pairs with separation distances less than 0.8 m to the number of
single muons to be $(7-8) \times 10^{-4}$, while our simulations
predict the ratio to be about $2.2 \times 10^{-4}$. We conclude that 
the enhancement of the decoherence function over conventional
muon bundle predictions seen by the MACRO experiment cannot be
explained by the effect of muon pair production alone, unless its cross-section
is underestimated by a factor of 3. Even in this case, it is difficult
to explain the difference in the shape of the pair separation distribution.

\vspace{0.8cm}
\noindent {\large \bf 4. Conclusions}
\vspace{0.5cm}

We have revised the process of muon pair production in rock using the
recently re-calculated cross-section for this process. Our three-dimensional 
Monte Carlo of muon propagation through rock shows that 
there is an observable excess (about $10\%$ at 3 km w.e.) in the number
of muon pairs with separation distances less than 1 m over conventional
muon bundles, even though the newly calculated cross-section is 2--5 times less
than that used before. Observation of this excess is within the reach
of existing large underground detectors capable of collecting hundreds of
thousands of multiple muon events.
However, the excess of muon pairs with separation distances less than 3 m,
recently published by the MACRO collaboration, cannot be explained by
muon pair production alone unless the cross-section of the process is 3
times larger than that used in this work.

The modified three-dimensional muon propagation code MUSIC, which includes
also muon pair production, can be obtained by request to 
v.kudryavtsev@sheffield.ac.uk

\pagebreak

\begin{table}[htb]
\caption{ Main characteristics of narrow muon bundles for different
depths in standard rock (columns 2-5) and water (column 6) 
(1 km w.e. = $10^5$ g/cm$^2$):
number of simulated muons $N_{sim}$,
number of single muons survived $N_{surv}$, 
ratio of the total number of double ($R_2$) triple ($R_3$) and 
double + triple ($R_b$) muon events
to that of single muons, 
mean energy of single muons $<E_{s}>$,
mean energy of muons in bundles $<E_{b}>$,
mean path of muon bundle in rock/water between the point of its production
and the observation level $<L>$,
weighted mean separation of muon pairs in double ($<D_2>$), triple
($<D_3>$), and double + triple ($<D_b>$) muon events,
weighted mean angular separation of muon pairs in bundles 
$<\alpha>$, and the ratio of
weighted number of pairs with separation distances less than 1 m to the number
of single muons $R_b(<1$ m). }

\vspace{1cm}
\begin{center}
\begin{tabular}{|c|c|c|c|c|c|}\hline
Depth, km w.e. &   1    &   3    &   5    &  10    & 3 km of water \\ \hline
$N_{sim}$  & $10^7$ & $10^7$ & $10^7$ & $10^7$ & $10^7$ \\ \hline
$N_{surv}$ & $4.01 \cdot 10^6$ & $3.41 \cdot 10^6$ & 
$1.94 \cdot 10^6$ & $4.12 \cdot 10^6$ & $3.47 \cdot 10^6$ \\ \hline
$R_2$ & $5.54 \cdot 10^{-5}$ & $1.81 \cdot 10^{-4}$ & 
$2.69 \cdot 10^{-4}$ & $4.03 \cdot 10^{-4}$ & $1.48 \cdot 10^{-4}$ \\ \hline
$R_3$ & $2.07 \cdot 10^{-5}$ & $7.21 \cdot 10^{-5}$ & 
$8.81 \cdot 10^{-5}$ & $1.04 \cdot 10^{-4}$ & $5.04 \cdot 10^{-5}$ \\ \hline
$R_b$ & $7.61 \cdot 10^{-5}$ & $2.53 \cdot 10^{-4}$ & 
$3.58 \cdot 10^{-4}$ & $5.08 \cdot 10^{-4}$ & $1.99 \cdot 10^{-4}$ \\ \hline
$<E_{s}>$, GeV &  $88$ & $247$ & $314$ & $366$ & $321$ \\ \hline
$<E_{b}>$, GeV & $221$ & $386$ & $545$ & $793$ & $513$ \\ \hline
$<L>$, km w.e. & $0.256$ & $0.585$ & $1.001$ & $2.032$ & $0.653$ \\ \hline
$<D_2>$, cm & $49$ & $81$ & $93$ & $110$ & $165$ \\ \hline
$<D_3>$, cm & $29$ & $33$ & $46$ & $69$ & $77$ \\ \hline
$<D_b>$, cm & $44$ & $68$ & $81$ & $101$ & $143$ \\ \hline
$<\alpha>$, deg & $1.11$ & $0.893$ & $0.757$ & $0.634$ & $0.615$ \\ \hline
$R_b(<1$ m) & $6.78 \cdot 10^{-5}$ & $1.94 \cdot 10^{-4}$ & 
$2.57 \cdot 10^{-4}$ & $3.21 \cdot 10^{-4}$ & $1.07 \cdot 10^{-4}$ \\ \hline
\end{tabular}
\end{center}
\end{table}

\pagebreak

\begin{figure}[htb]
\begin{center}
\epsfig{figure=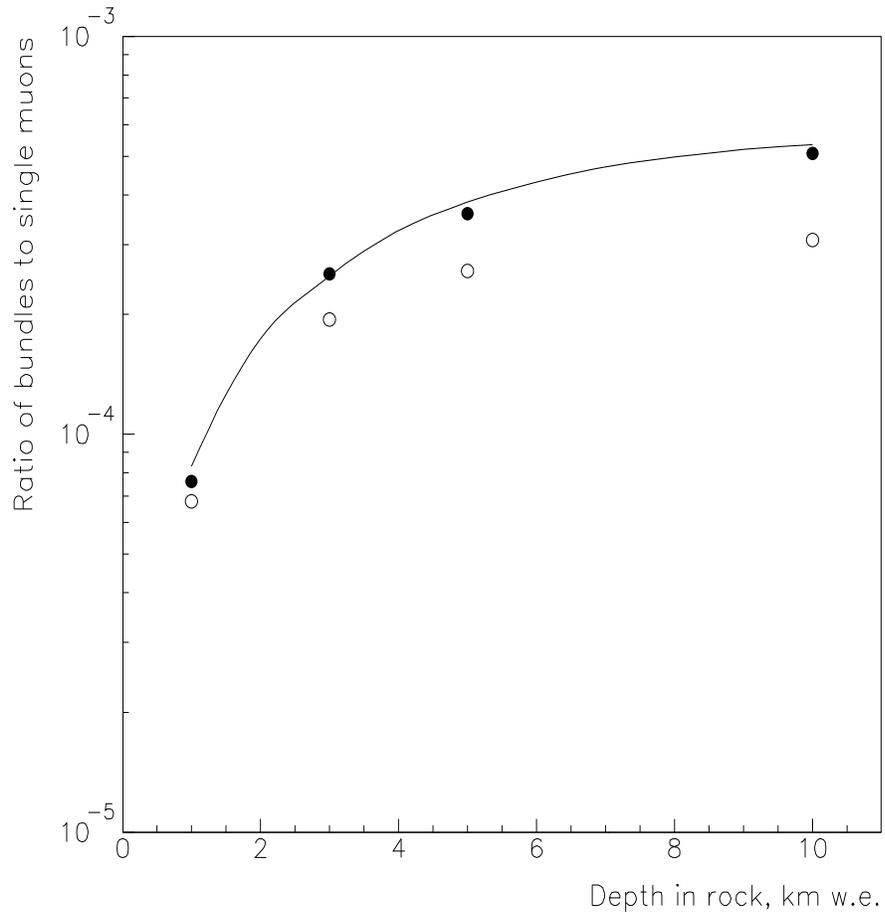,width=12cm,height=15cm}
\caption{
Ratios of narrow muon bundles to single muons {\it vs} depth in standard rock:
{\it filled circles} -- whole sample of muon bundles;
{\it open circles} -- weighted pairs with distances less than 1 m between muons;
{\it solid curve} -- results of \cite{KKP} for the whole sample of 
muon bundles.} 
\end{center}
\end{figure}

\pagebreak

\begin{figure}[htb]
\begin{center}
\epsfig{figure=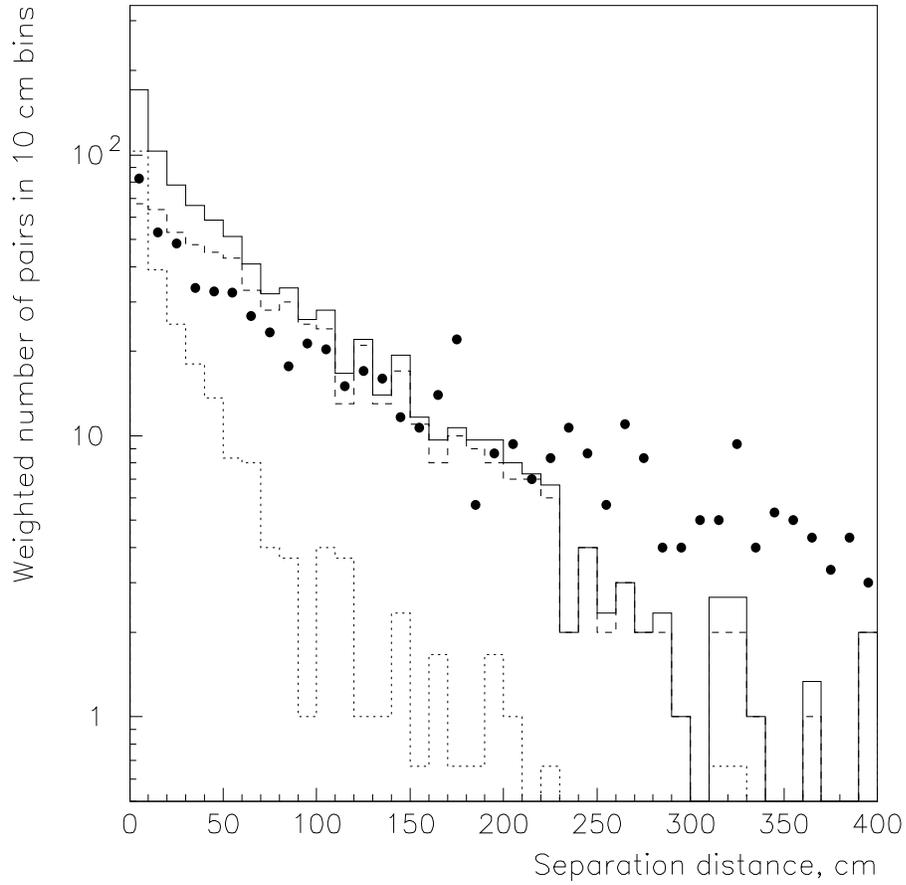,width=12cm,height=15cm}
\caption{
Weighted distributions of separation of muon pairs
in double (dashed histogram), triple (dotted histogram) and all 
(double + triple)
(solid histogram) muon events at 3 km w.e. in rock and 
the distribution for all muon bundles at 3 km depth in water
(filled circles).}
\end{center}
\end{figure}

\end{document}